\documentclass[12pt,english]{iopart}
\usepackage[T1]{fontenc}
\usepackage[latin9]{inputenc}
\usepackage[active]{srcltx}
\usepackage{array}
\usepackage{amsbsy}
\usepackage{graphicx}

\makeatletter

\providecommand{\tabularnewline}{\\}

\usepackage{iopams}
\usepackage{setstack}

\pdfoutput=1

\usepackage{amsbsy}\usepackage{array}

\providecommand{\tabularnewline}{\\}

\usepackage{iopams}\usepackage{setstack}

\usepackage{babel}

\usepackage{babel}

\makeatother

\usepackage{babel}
\begin{document}

\title{The fluctuating $\alpha$-effect and Waldmeier relations in the nonlinear
dynamo models }

\author{V.V. Pipin$^{1-3}$ and D.D. Sokoloff$^{3,4}$}

\address{$^{1}$Institute of Solar-Terrestrial Physics, Russian Academy of
Sciences,\\
 $^{2}$ Institute of Geophysics and Planetary Physics, UCLA, Los
Angeles, CA 90065, USA \\
 $^{3}$NORDITA, Roslagstullsbacken 23, 106 91 Stockholm, Sweden\\
 $^{4}$Department of Physics, Moscow State University, Moscow, 119991, Russia}
\begin{abstract}
We study the possibility to reproduce the statistical relations of
the sunspot activity cycle, like the so-called Waldmeier relations,
the cycle period - amplitude and the cycle rise rate - amplitude relations,
by means of the mean field dynamo models with the fluctuating $\alpha$-effect.
The dynamo model includes the long-term fluctuations of the $\alpha$-effect
and two types of the nonlinear feedback of the mean-field on the $\alpha$-effect
including the algebraic quenching and the dynamic quenching due to
the magnetic helicity generation. We found that the models are able
to reproduce qualitatively and quantitatively the inclination and dispersion across the
Waldmeier relations with the 20\% fluctuations of the $\alpha$-effect.
The models with the dynamic quenching are in a better agreement with
observations than the models with the algebraic $\alpha$-quenching.
We compare the statistical distributions of the modeled parameters,
like the amplitude, period, the rise and decay rates of the sunspot
cycles, with observations.
\end{abstract}
\maketitle

\section{Introduction}

It is observed that the sunspot's activity is organized in time and
latitude and forms the large scales patterns which are called the
Maunder butterfly diagram. This pattern is believed to be produced
by the large-scale toroidal magnetic field generated in the convection
zone. Another component of the solar activity is represented by the
global poloidal magnetic field extending outside the Sun and shaping
the solar corona. Both components synchronously evolve as the solar
11-year cycle progresses. The global poloidal field reverses the sign
in the polar regions near the time of maximum of sunspot activity.

A remarkable feature of cyclic solar activity is that it is far to
be just a cycle. Cycle amplitude and shape varies from one cycle to
the other and prognostic abilities of any study of solar activity
looks as its very attractive destination. Solar activity
observations give various hints that various tracers of solar
activity which are exploited to quantify the phenomenon demonstrate
some relation one to the other what opens a possibility to predict
future evolution of solar activity basing on available observations
of other indices. Waldmeier \cite{w35} pointed out at first this
option (an inverse correlation between the length of the ascending
phase of a cycle, or its \textquotedbl{}rise time\textquotedbl{},
and the peak sunspot number of that cycle) and applied it,
\cite{w36}, to give a prediction for the following cycle. The latter
paper is in practice the first accessible (at least for German
speaking readers) paper in the area. Later other relation of this
this type was suggested and summarized as Waldmeier relations. This
development was clearly summarized by \cite{vetal86} and recently by
\cite{hathaw02}. The nature of the physical processes, that are
manifested in the Waldmeier relations, is not clear, see discussion,
e.g., in \cite{camschu07,detal08,karak11}. It seems to be
remarkable, however, that these statistical properties of magnetic
activity are also existed for the other tracers related with the
sunspot activity (e.g., sunspot group and squares of sunspot groups,
see \cite{vetal86,hathaw02,karak11}), and even for the other kind of
the solar and stellar activity indices, e.g., for the Ca II index
\cite{soon94}. The Waldmeier relations are considered as a valuable
test of the dynamo models \cite{kitiash09,karak11,pk11}.

A natural way to push the understanding of the problem forward is
to clarify the physics underlying Waldmeier relations. It is more
or less accepted that cyclic solar activity is driven by a dynamo,
i.e. a mechanism which transforms kinetic energy of hydrodynamical
motions into magnetic one. Most of the current solar dynamo models
suggest that the toroidal magnetic field that emerges on the surface
and forms sunspots is generated near the bottom of the convection
zone, in the tachocline or just beneath it in a convection overshoot
layer (see, e.g., \cite{stix:02}). This kind of dynamo can be approximated
by the Parker's surface dynamo waves \cite{park93}. The direction
of the dynamo waves propagation is defined by the Parker-Yoshimura
rule \cite{yosh1975}. It states that for the $\alpha\Omega$ kind
dynamo the waves propagates along iso-surfaces of the angular velocity.
The propagation process can be modified by the turbulent transport
(associated with the mean drift of magnetic activity in the turbulent
media by means turbulent mechanisms), by the anisotropic turbulent
diffusivity (see, \cite{k02}), and by meridional circulation \cite{choud95}.
A viewpoint, which is an alternative to the Parker's surface dynamo
waves is presented by the distributed dynamo with subsurface shear,
e.g. \cite{b05}. The dynamo waves here propagates along the radius
in the main part of the solar convection zone, \cite{k02}. The near
surface activity is shaped by the subsurface shear. One more option
is the flux-transport dynamo, e.g. \cite{choud95,dc99}.

In the context of dynamo theory, the Waldmeier relations have to be
explained by some mechanism which varies amplitude and shape of activity
cycle and fluctuations $\alpha$-effect are considered below as such
mechanism. This idea extend the approach proposed in  \cite{pk11} to
explain these relations by changing the magnitude of the
$\alpha$-effect. 

The physical idea underlying this mechanism can be presented as follows.
$\alpha$-coefficient is a mean quantity taken over ensemble of convective
vortexes. Number $N$ of the vortexes in solar convective shell is
large however much smaller then, say, the Avogardo number, so fluctuations
being proportional to $N^{-1/2}$ may be not negligible. Particular
choice of $N$ is obviously model dependent however if we take just
for orientation $N=10^{4}$ then $N^{-1/2}=0.01$. Taking into account
that $\alpha$ is usually about 1/10 of turbulent velocity we consider
a dozen percent of $\alpha$-fluctuations as a comfortable estimate.
From the other hand, governing equations for large-scale solar magnetic
field deal with spatial averaging and have to include a contribution
of $\alpha$-fluctuations, \cite{h93}.

A straightforward application of the idea with vortex turnover time
and vortex size as correlation time and length for $\alpha$-fluctuations
needs fluctuations much larger then mean $\alpha$. \cite{moss-sok08},
\cite{uetal09} based on experiences in direct numerical simulations,
e.g. \cite{bs02}, and results of current helicity (related to $\alpha$)
observation in solar active regions, e.g. \cite{zetal10} considered
$\alpha$-fluctuations with correlation time comparable with cycle
length and correlation length comparable with the extent of the latitudinal
belts where sunspots occur to conclude that a reasonable $\alpha$-noise
of order of few dozen percents is sufficient to explain Grand minima
of solar activity. The aim of this paper is to apply this idea to
explain Waldmeier relations.

\section{Basic equations}

\subsection{2D model}

The dynamo model is based on the standard mean-field induction equation
in perfectly conductive media \cite{krarad80}:
\[
\frac{\partial\mathbf{B}}{\partial t}=\boldsymbol{\nabla}\times\left(\mathbf{\boldsymbol{\mathcal{E}}+}\mathbf{U}\times\mathbf{B}\right)
\]
 where $\boldsymbol{\mathcal{E}}=\overline{\mathbf{u\times b}}$ is
the mean electromotive force, with $\mathbf{u,\, b}$ being the turbulent
fluctuating velocity and magnetic field respectively; $\mathbf{U}$
is the mean velocity (differential rotation). The axisymmetric magnetic
filed:
\[
\mathbf{B}=\mathbf{e}_{\phi}B+\nabla\times\frac{A\mathbf{e}_{\phi}}{r\sin\theta}
\]
 $\theta$ - polar angle. We have used the expression for $\boldsymbol{\mathcal{E}}$
obtained by \cite{pi08Gafd} (hereafter P08) and write it as follows:
\begin{equation}
\mathcal{E}_{i}=\left(\alpha_{ij}+\gamma_{ij}\right)\overline{B}_j-\eta_{ijk}\nabla_{j}\overline{B}_{k}.\label{eq:EMF-1}
\end{equation}
 Tensor $\alpha_{i,j}$ represents the alpha effect, including the
hydrodynamic and magnetic helicity contributions,
\begin{equation}
\alpha_{ij}=C_{\alpha}\left(1+\xi\right)\psi_{\alpha}(\beta)\sin^{2}\theta\alpha_{ij}^{(H)}+\alpha_{ij}^{(M)},\label{alp2d}
\end{equation}
 where the hydrodynamical part of the $\alpha$-effect, $\alpha_{ij}^{(H)}$,
$\xi$ is the noise, and the quenching function, $\psi_{\alpha}$, are
given in Appendix (see also in \cite{pk11apjl}). The hydrodynamic
$\alpha$-effect term is multiplied by $\sin^{2}\theta$ ($\theta$ is
co-latitude) to prevent the turbulent generation of magnetic field
at the poles. The contribution of the small-scale magnetic helicity
$\overline{\chi}=\overline{\mathbf{a\cdot}\mathbf{b}}$ ($\mathbf{a}$
is a fluctuating vector-potential of magnetic field) to the
$\alpha$-effect is defined as
$\alpha_{ij}^{(M)}=C_{ij}^{(\chi)}\overline{\chi}$, where
coefficient $C_{ij}^{(\chi)}$ depends on the turbulent properties
and rotation, and is given in Appendix. The other parts of
Eq.(\ref{eq:EMF-1}) represent the effects of turbulent pumping,
$\gamma_{ij}$, and turbulent diffusion, $\eta_{ijk}$. They are the
same as in PK11. We describe them in Appendix.
\begin{figure}
\begin{centering}
\includegraphics[width=0.9\textwidth]{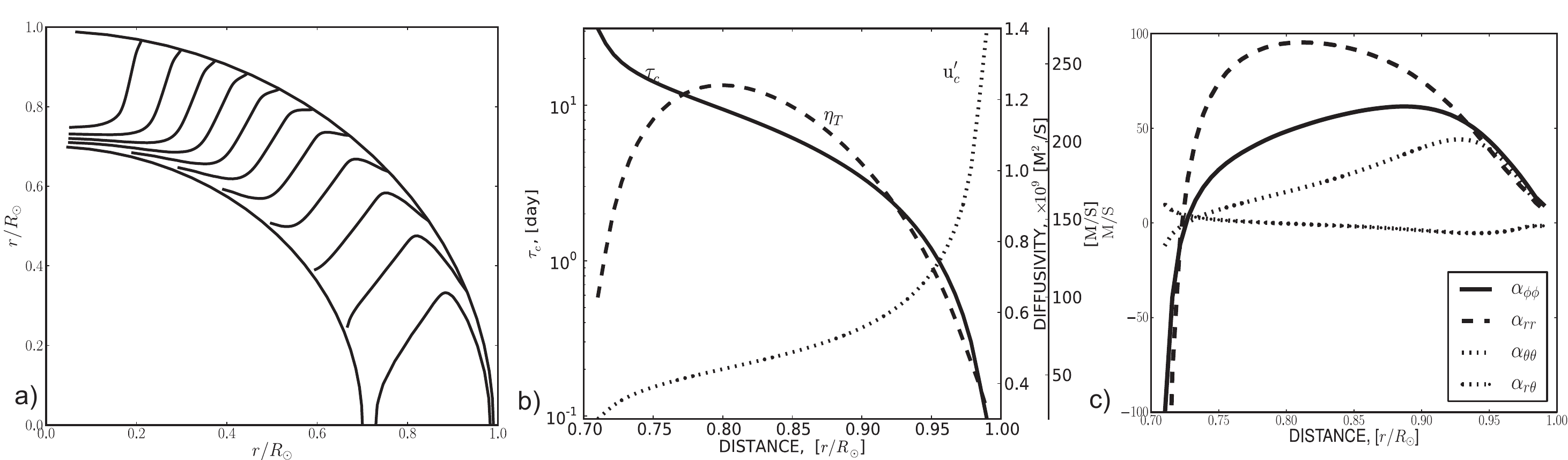}
\par\end{centering}

\caption{\label{fig1-1}Parameters of the solar convection zone: a) the contours
of the constant angular velocity plotted for the levels $(0.75-1.05)\Omega_{0}$
with a step of $0.025\Omega_{0}$, $\Omega_{0}=2.86\cdot10^{-7}s^{-1}$;
b) turnover convection time $\tau_{c}$, and the RMS convective velocity
$u'_{c}$ and the background turbulent diffusivity $\eta_{T}^{(0)}$
profiles; c) the radial profiles of the $\alpha$-effect tensor components.}
\end{figure}

The nonlinear feedback of the large-scale magnetic field to the $\alpha$-effect
is described as a combination of an ``algebraic'' quenching by function
$\psi_{\alpha}\left(\beta\right)$ (see Appendix and \cite{pk11}),
and a dynamical quenching due to the magnetic helicity conservation
constraint. The magnetic helicity, $\overline{\chi}$ , subject to
a conservation law, is described by the following equation \cite{kle-rog99,bra-sub:04d,sub-bra:04}:
\begin{eqnarray}
\frac{\partial\overline{\chi}}{\partial t} & = & -2\left(\boldsymbol{\mathcal{E}\cdot}
\overline{\mathbf{B}}\right)-\frac{\overline{\chi}}{R_{\chi}\tau_{c}}+\boldsymbol{\nabla}
\cdot\left(\eta_{\chi}\boldsymbol{\nabla}\bar{\chi}\right),\label{eq:hel}
\end{eqnarray}
 where $\tau_{c}$ is a typical convection turnover time. Parameter
$R_{\chi}$ controls the helicity dissipation rate without specifying
the nature of the loss. It seems to be reasonable that the helicity
dissipation is most efficient in the near surface layers because of
the strong decrease of $\tau_{c}$ (see Figure 1b). The last term in
Eq.(\ref{eq:hel}) describes the diffusive flux of magnetic helicity
\cite{mitra10}. We use the solar convection zone model computed by
\cite{stix:02}, in which the mixing-length is defined as
$\ell=\alpha_{MLT}\left|\Lambda^{(p)}\right|^{-1}$, where
$\mathbf{\boldsymbol{\Lambda}}^{(p)}=\boldsymbol{\nabla}\log\overline{p}\,$
is the pressure variation scale, and $\alpha_{MLT}=2$. The turbulent
diffusivity is parametrized in the form,
$\eta_{T}=C_{\eta}\eta_{T}^{(0)}$, where
$\eta_{T}^{(0)}={\displaystyle \frac{u'\ell}{3}}$ is the
characteristic mixing-length turbulent diffusivity, $\ell$ and $u'$
are the typical correlation length and RMS convective velocity of
turbulent flows, respectively and $C_{\eta}$ is a constant to
control the intensity of turbulent mixing. In the paper we use
$C_{\eta}=0.05$. The differential rotation profile,
$\Omega=\Omega_{0}f_{\Omega}\left(x,\mu\right)$, $x=r/R_{\odot}$,
$\mu=\cos\theta$ is a modified version of an analytical
approximation to helioseismology data, proposed by \cite{antia98},
see Fig.~1a.
\begin{figure}
\includegraphics[width=0.8\paperwidth]{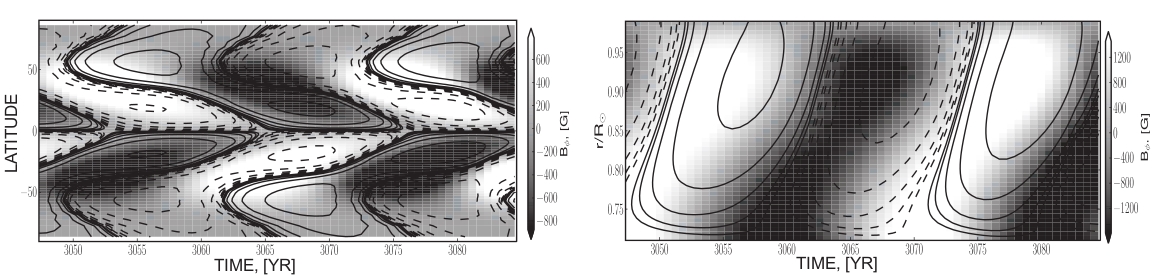} \caption{\label{2dbat}The typical time-latitude and the time-radius (at the
$30^{\circ}$ latitude) diagrams of the toroidal field (grey scale),
the radial field (contours at left panel) and the poloidal magnetic
field (contours at the right panel) evolution in 2D1 model (see Table
1). The toroidal field averaged over over the subsurface layers in
the range of $0.9-0.99R_{\odot}$ , the radial field is taken at the
top of the convection zone. }
\end{figure}

We use the standard boundary conditions to match the potential field
outside and the perfect conductivity at the bottom boundary. As discussed
above, the penetration of the toroidal magnetic field in to the near
surface layers is controlled by the turbulent diffusivity and pumping
effect. For magnetic helicity, similar to \cite{guero10} and \cite{mitra11},
we use the time dependent conditions provided be Eq.\ref{eq:hel} and
the helicity flux conservation the condition
$\boldsymbol{\nabla_{r}}\bar{\chi}=0$ is applied  at the bottom and at
the top of domain. The latter gives a smooth transfer for solutions
with and without the diffusive helicity flux.

The left panel on the Fig.~\ref{2dbat} shows the typical the
time-latitude diagram for the toroidal magnetic averaged over the
subsurface layers $0.9-0.99R_{\odot}$ and the radial magnetic at the
top of the integration domain. The right panel shows the time-radius
the time-radius diagram for the toroidal an poloidal magnetic field
evolution at $30^{\circ}$ latitude.

We demonstrate it by Fig.~3 which shows the time-latitude diagrams
for toroidal and radial magnetic field evolution for the models 1D1,
1D3 and 2D1. For the latter model we show the toroidal magnetic
averaged over the subsurface layers $0.9-0.99R_{\odot}$ and the
radial magnetic field is given for the top of the integration
domain. For the model 2D1 we show the time-radius diagram for the
toroidal an poloidal magnetic field evolution at $30^{\circ}$
latitude. The other models listed in Table 1, having the same
general patterns of the magnetic field evolution, are differed from
the models shown on the Fig.~3 in some details (mostly associated
with magnetic helicity evolution).

\subsection{1D model}

For comparison with the previous studies and also to study how the
additional dimension affect the statistical properties of the dynamo
we consider the $1D$ model similar to that studied by \cite{moss-sok08}:

\begin{eqnarray}
\frac{\partial A}{\partial t} & = & \sin\theta\left(\left(1+\xi\right)\cos\theta
\psi_{\alpha}(B)+\chi\right)B+\sin\theta\frac{\partial}{\partial\theta}\left(\frac{1}
{\sin\theta}\frac{\partial A}{\partial\theta}\right)-\eta_{CZ}A,\label{eq:D1a}\\
\frac{\partial B}{\partial t} & = & -\mathcal{D}\tilde{\Omega}\left(\theta\right)
\frac{\partial A}{\partial\theta}+\frac{\partial}{\partial\theta}\left(\frac{1}
{\sin\theta}\frac{\partial\sin\theta B}{\partial\theta}\right)-\eta_{CZ}B,\label{eq:D1sys}
\end{eqnarray}
 where the large-scale radial shear $\tilde{\Omega}\left(\theta\right)=\partial\Omega/\partial r$.
The $1D$ model employs two possibilities for the shear profile. In
one case we put $\tilde{\Omega}\left(\theta\right)=1$, that give
us the model explored by \cite{moss-sok08}. In another case we use
\begin{equation}
\tilde{\Omega}=\frac{1}{10}\left(5\sin^{2}\theta-4\right),\label{eq:shear}
\end{equation}
 which is suggested by \cite{kit-rud-kuk}. In agreement with the
helioseismology results for the bottom of the convection zone, this
profile is positive in equatorial regions and negative near the poles.
The magnetic field strength in Eq.(\ref{eq:D1sys}) is measured in
the units of the equipartition magnetic field strength and the time
is normalized to the typical diffusive time, $R_{\odot}^{2}/\eta_{T}^{(0)}$.
The evolution of the magnetic helicity for the 1D model is governed
by equation:
\begin{eqnarray}
\frac{\partial\overline{\chi}}{\partial t} & = & -2\left((1+\xi)\cos\theta
\psi_{\alpha}(B)+\chi\right)B^{2}-2B\frac{\partial}{\partial\theta}\left(\frac{1}
{\sin\theta}\frac{\partial A}{\partial\theta}\right)\label{eq:1dhel}\\
 & + & \frac{2}{\sin^{2}\theta}\frac{\partial A}{\partial\theta}\frac{\partial
 \sin\theta B}{\partial\theta}-\frac{\overline{\chi}}{R_{\chi}}+\frac{\eta_{\chi}}
 {\sin\theta}\frac{\partial}{\partial\theta}\left(\frac{1}{\sin\theta}\frac{\partial\bar{\chi}}{\partial\theta}\right).\nonumber
\end{eqnarray}

In what follows we will discuss the 1D models with the constant shear,
because they are more relevant to compare with observations. The differences
in results for the 1D models with the variable shear given by Eq.(\ref{eq:shear})
will be briefly mentioned in subsequent sections.
\begin{figure}
\includegraphics[width=0.9\paperwidth]{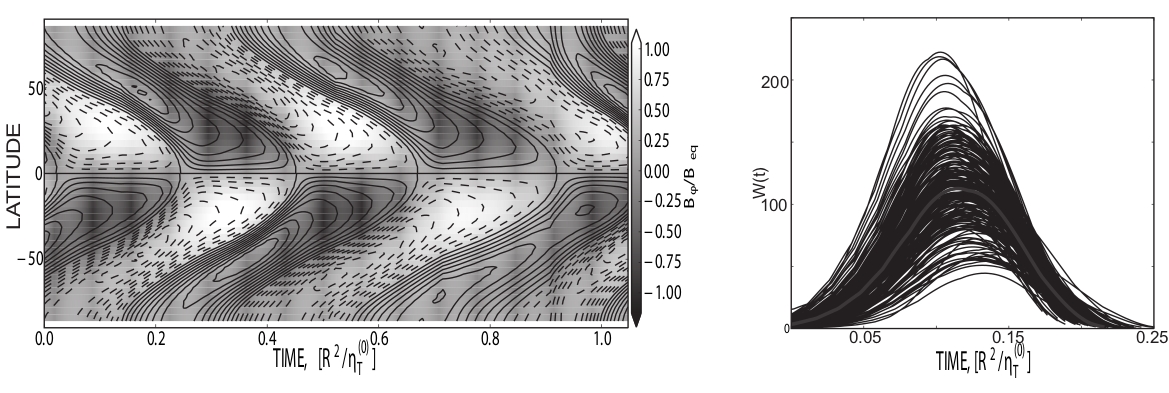}

\caption{\label{fig:1d}The left panel shows the time-latitude diagrams of
the toroidal field (grey scale) and the radial field (contours) for
the 1D1 model (see Table 1). The right panel shows the estimated sunspot
number in the the separated cycles in the 1D1 model (see Section \ref{ssndef}). }
\end{figure}

Summarizing, we exploit much more detailed and realistic dynamo models
then \cite{moss-sok08}, \cite{uetal09}. Our point is that Waldmeier
relations are a much more delicate phenomena rather Grand minima and
the bulk of our knowledge concerning recent solar cycles is much more
rich then that one for remote past when Grand minima took place.

\subsection{Noise model}

The noise, $\xi$, contributes in the hydrodynamic part of the
$\alpha$-effect (see, Eqs.(\ref{alp2d},\ref{eq:D1a})). Following to
\cite{uetal09} the models employ the long-term Gaussian fluctuating
$\xi$ of the small amplitude with RMS deviation given in the Table 1
(last column). The time of the renewal of the $\xi$ is equal to the
period of the model. The random numbers were generated with help of
the standard F90 subroutine quality of contemporary standard noise
generator subroutine is shown to be sufficient for such kind of
modelling, see e.g. \cite{as06}). It would be more realistic to
consider the renewal time as the fluctuating quantity as well, but
we would like to separate this effect for the different study. Also,
we found that the models which employ the magnetic helicity effect
show the very intermittent long term behaviour. This makes the
analysis procedure (e.g., division to subsequent cycles) more
complicated. We isolate ourselves from these phenomena by
considering the noise models with the lower RMS in case if the
magnetic helicity is employed. \footnote{In part, the given problem
is likely due to the very rough model for the Wolf number, see
Eq.(\ref{eq:wolf}).}
\begin{table}
\begin{centering}
\begin{tabular}{|c|c|c|c|c|c|c|c|}
\hline
Model  & $\eta_{CZ}$  & $\overline{\chi}$  & $\eta_{\chi}/\eta_{T}$  & $R_{\chi}$  & $B_{0}$  & $C_{W}$  & $\sigma$\tabularnewline
\hline
1D1  & 1  & 0  & 0  & 10  & 3  & 1200  & 0.15\tabularnewline
2D1  &  & Eq.(2)  & $10^{-5}$  & 200  & 800  & 1  & 0.15\tabularnewline
2D2  &  & Eq.(2)  & $0.3$  & $10^{6}$  & 200  & 1  & 0.15\tabularnewline
\hline
\end{tabular}
\par\end{centering}

\caption{Parameters the dynamo models: the type of the nonlinear
quenching of the $\alpha$-effect, if the magnetic helicity is
$\overline{\chi}=0$ then the model employ only the algebraic
quenching which is described by $\psi_{\alpha}$and otherwise by the
dynamic quenching due to magnetic helicity described by
Eq.(\ref{eq:hel}) or Eq.(\ref{eq:1dhel}); $\eta_{\chi}/\eta_{T}$ is
the ratio between the turbulent magnetic helicity diffusivity and
the turbulent magnetic diffusivities; the profile of the shear in
the 1D models; the $\alpha$-effect parameter in the 2D models; the
parameter $R_{\chi}$ controls the helicity dissipation rate; the
parameter $B_{0}/B_{{\rm eq}}$ controls the sunspot number parameter
in the 1D models. It is the ratio between the typical strength of
the toroidal magnetic field producing the sunspots and the
equipartition magnetic field strength; $B_{0}$ is the typical
strength of the toroidal magnetic field controlling the sunspots
number parameter in the 2D models; $C_{W}$ is the parameter to
calibrate the modeled sunspot number relative to observations;
$\sigma$ is the standard deviation of the Gaussian noise in the
model}
\end{table}

\subsection{The sunspot cycle model and the Waldmeier relations\label{ssndef}}

In the paper we define the Waldmeier relations as the set of the
mean properties of the sunspot cycle. We will deal with the
following properties of the Wolf sunspot number (which is taken
either from observational database or simulated from the model): the
relation between period and amplitude of the same cycle, the
relation between rise rate and amplitude of the cycle and the shape
of the sunspot cycle, characterized by the ratio between the decay
rate and the rise rate in the cycle. The other kind of relations,
like the link between the rise time and amplitude of the cycle, can
be considered as the derivative from the above relation. For
comparison with other analysis of the observational data and also
with the results of the dynamo models presented by Karak and
Choudhuri\cite{karak11} we show the results for the rise time of the
cycles as well, the relation between the rise time and amplitude of
the cycle and the relation between the cycle amplitude and period of
the preceding cycle (see, \cite{vetal86} and \cite{hathaw02}). The
amplitude of the cycles is defined by difference between the maximum
sunspot number and the sunspot number in the preceding minimum. {
Even for the harmonic cycles the latter
differs from zero due to the spatial overlap in subsequent cycles.} The
period of the cycle is equal to the time between the subsequent
minima, the rise time of the cycle is defined by the difference
between the moment of the cycle maximum and the moment of the
preceding minimum of the cycle. The rise rate is defined as the
ratio between the difference of the sunspot number amplitude during
maximum and minimum of the cycle and the rise time of the cycle. The
similar definition is for the decay rate of the cycle.

Remind that sunspots are not directly presented in dynamo models and
we have to relate its number to a quantity involved in a dynamo model
under consideration. We assume that the sunspots are produced from
the toroidal magnetic fields by means of the nonlinear instability
and avoid to consider the instability in details. To model the sunspot
number $W$ produced by the dynamo we use the following anzatz
\begin{equation}
W\left(t\right)=C_{W}\left\langle B_{{\rm max}}\right\rangle _{SL}
\exp\left(-\frac{B_{0}}{\left\langle B_{{\rm max}}\right\rangle _{SL}}\right),\label{eq:wolf}
\end{equation}
 where for the 2D models $\left\langle B_{{\rm max}}\right\rangle _{SL}$
is the maximum of the toroidal magnetic field strength over latitudes
averaged over the subsurface layers in the range of $0.9-0.99R_{\odot}$
and for the 1D models $\left\langle B_{{\rm max}}\right\rangle _{SL}$
is simply the maximum of the toroidal magnetic field strength over
latitudes; $B_{0}$ is the typical strength of the toroidal magnetic
field that is enough to produce the sunspot; $C_{W}$ is the parameter
to calibrate the modeled sunspot number relative to observations.
The all parameters which were employed in the different models are
listed in the Table 1.

{ In the dynamo models we explore the effect of the Gaussian
fluctuations of the $\alpha$-effect, or  parameter $C_{\alpha}$
with} the typical time equal to the period of the cycle and the
standard deviations less than $0.2C_{\alpha}$. In the models presented
here we fix  the
standard deviation to $0.15C_{\alpha}$.

\begin{table}
\begin{tabular}{|>{\centering}p{2.5cm}|>{\centering}p{2.5cm}|>{\centering}p{2.5cm}|>{\centering}p{2.5cm}|>{\centering}p{2.5cm}|>{\centering}p{2.6cm}|}
\hline
 & 1D1  & 2D1  & 2D2  & SIDC  & NIMV

(2004)\tabularnewline
\hline
\hline
Period  & 11.02$\pm$0.66  & 11.07$\pm$1.08  & 10.97$\pm$0.92  & 11.01$\pm$1.12  & 11.02$\pm$1.49\tabularnewline
\hline
Amplitude  & 115.7$\pm$33.6  & 103.3$\pm$40.5  & 96.3$\pm$25.7  & 108.2$\pm$38.1  & 87.6$\pm$43.9\tabularnewline
\hline
Rise Rate  & 18.62$\pm$6.14  & 25.39$\pm$11.95  & 19.91$\pm$5.95  & 25.81$\pm$12.74  & 19.48$\pm$13.38 \tabularnewline
\hline
Rise Time & 6.11$\pm$.33 & 4.06$\pm$.77 & 4.73$\pm$.36 & 4.32$\pm$1.07 & 4.82$\pm$1.32\tabularnewline
\hline
Shape  & 1.27$\pm$0.2  & .59$\pm$0.15  & .77$\pm$0.08  & .69$\pm$0.31  & .83$\pm$0.34 \tabularnewline
\hline
Rise Rate - Amplitude  & 5.4x+14.2$\pm$3.0

0.99 & 3.3x+18.8$\pm$7.6

0.98 & 4.2x+12.4$\pm$5.6

0.98 & 2.9x+33.2$\pm$8.9

0.97 & 3.1x+27.8$\pm$15.7

0.93\tabularnewline
\hline
Period - Amplitude(a)  & -31.7x+463.9

$\pm$26.2

-0.63 & -17.5x+298.0

$\pm$34.0

-0.54 & -17.25x+2856

$\pm$20.3

-0.62 & -23.6x+368.5

$\pm$28.0

-0.68 & -8.4x+179.9

$\pm$42.0

-0.29\tabularnewline
\hline
Period - Amplitude(b)  & -17.9x+312.3

$\pm$31.4

-0.35 & -8.9x+202.9

$\pm$38.9

-0.28 & -6.3x+165.4

$\pm$25.0

-0.22 & -11.2x+231.7

$\pm$35.9

-0.33 & -6.9x+163.4

$\pm$42.7

-0.23\tabularnewline
\hline
Rise Time - Amplitude  & -82.1x+617.4

$\pm$18.3

-0.84 & -25.6x+207.5

$\pm$35.3

-0.49 & -33.0x+252.8

$\pm$22.7

-0.47 & -26.7x+234.

$\pm$24.

-0.75 & -16.1x+165.4

$\pm$38.5

-0.48\tabularnewline
\hline
Rise Rate - Decay Rate  & 1.0x+4.0$\pm$3.1

0.9 & 0.43x+3.3$\pm$2.2

0.92 & 0.68x+1.6$\pm$1.6

0.93 & 0.34x+6.4$\pm$2.6

0.85 & 0.42x+5.3$\pm$4.1

0.81\tabularnewline
\hline
\end{tabular}

\caption{First five rows contain information for the mean and variance (standard
deviation) for the parameters of the sunspot cycles in the different
data set. The shape of the cycle is defined as ratio between the decay
rate and the rise rate of the cycle. Last five rows show the linear
fits with the mean-square error bar and the correlation coefficient.
In the relation Period-Amplitude (a) we compare the cycle amplitudes
to period of the \emph{preceding} cycle (see \cite{hathaw02,vetal86}),
and in the relation Period-Amplitude (b) we compare these parameters
for the \emph{same} cycle.}

\end{table}

\begin{figure}
\includegraphics[width=0.9\textwidth]{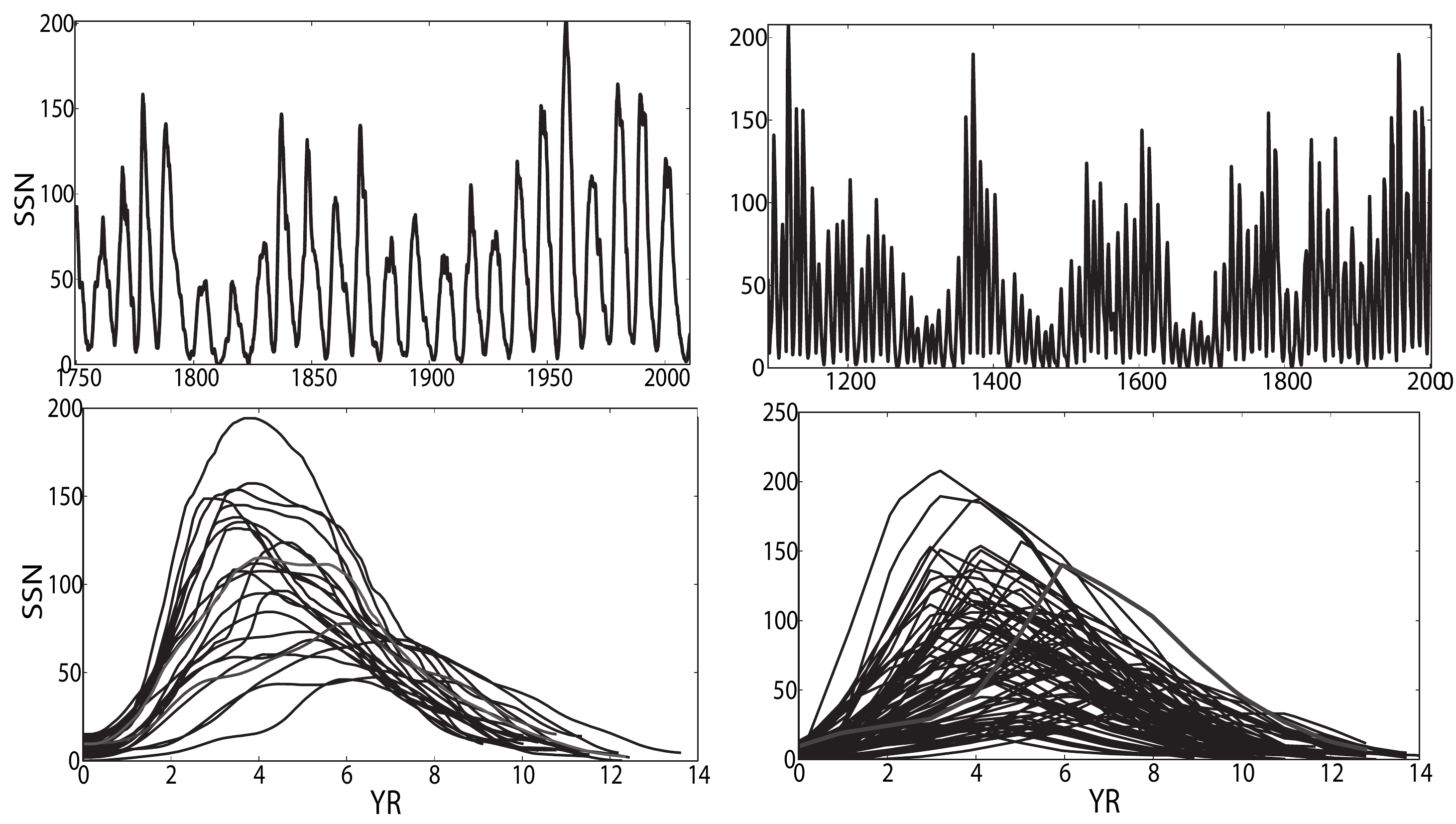} \caption{\label{fig:The-sunspot-data}The sunspot data sets. Upper raw: left
- SIDC and right - \cite{nag04}(NIMV), lower raw - corresponding
cycles distributions}
\end{figure}

For comparison with simulation we use the smoothed data set from \cite{sidc}
which starts at 1750. Choosing this data set we appreciate that in
principle Waldmeier relations can be valid for normal cycle only and
their applicability to epochs of Grand minima of solar activity must
be addressed separately. Available instrumental data concerning solar
activity in XVII - early XVIII centuries gives a limited possibility
only to address this important point which obviously is out of the
scope of this paper. From the other hand, there are various indirect
(mainly isotopic) tracers of solar activity which give a limited information
concerning its shape over much longer time interval rather instrumental
data. Our point is that Waldmeier relations and the regularities of such long-term
time series (see, e.g.,\cite{mordv99,mordv10}) have to be discussed in a separate paper and here use
as an illustrative example the extended time series of the sunspot
data proposed by \cite{nag04} (referred hereafter as NIMV). %
These data sets are shown on Fig.~\ref{fig:The-sunspot-data}. The
Table 2 contains the linear fits and correlations between the
different parameters of the cycles for observational data sets and
for the dynamo models as well. In particular, the parameters of the
relation between rise time and amplitude and parameters of the
Amplitude-Period effect (a) and (b) (associated with period of the
\emph{preceding} and the \emph{same} cycle) for SIDC data set are in
a good agreement with the results by Vitinskij et al \cite{vetal86}
and Hathaway et al\cite{hathaw02}. The similar conclusion can be
done if we compare our analysis for SIDC data set for the relation
between rise rate and amplitude of the cycles with analysis given by
Vitinskij et al \cite{vetal86}.

\section{Results}

\begin{figure}[t]
 \includegraphics[width=0.9\textwidth]{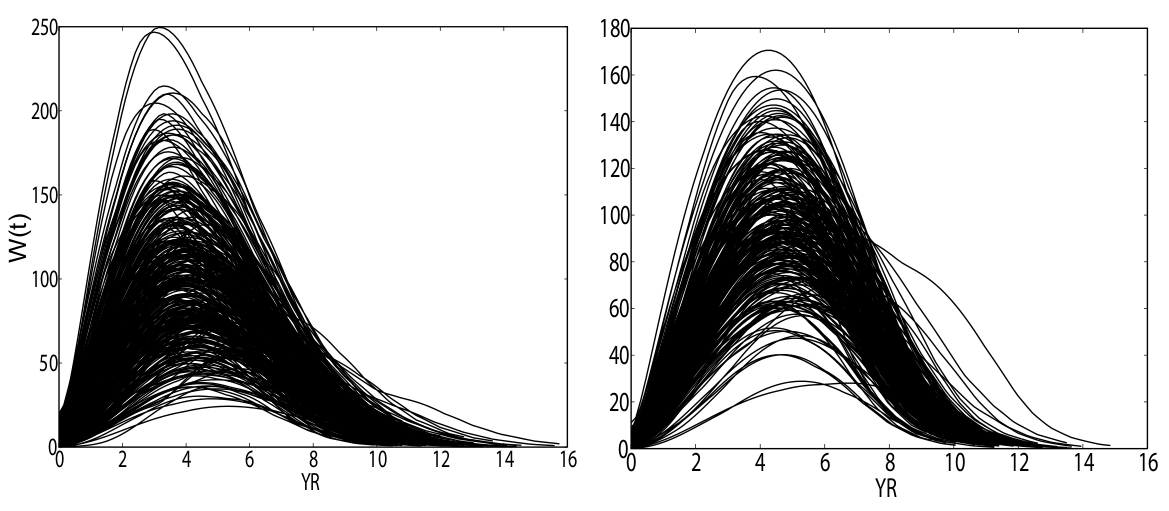} \caption{\label{fig:2d}Left panel shows cycles distributions for the model
2D1 and the right panel - model 2D2.}
\end{figure}

The typical time-latitude diagrams for the dynamo models were shown
in Figures \ref{2dbat} and \ref{fig:1d}. The shape of the simulated
sunspot cycles in 1D1 model can be seen on the right panel Figure
\ref{fig:1d}. The simulated sunspot cycles for the 2D1 and 2D2 models
are shown on the the Figure \ref{fig:2d}. We can conclude that the
shape of the simulated sunspot cycles (and, perhaps, the associated
Waldmeier relations) is directly related with the spatial shape of
the toroidal magnetic field evaluational patterns. For example, in
the 1D1 model the maximum of the butterfly diagram is very close to
equator and butterfly wing is elongated toward the pole. In such a
pattern of the toroidal magnetic field evolution the decay phase of
the sunspot activity is shorter than the rise phase. The opposite
situation is in the models 2D1 and 2D2. The physical mechanisms which
produce the short rise and the long decay of the toroidal magnetic
field activity were discussed recently by Pipin and Kosovichev \cite{pk11}.

To proceed further we would like to discuss the statistical properties
of the cycle parameters those involved in the Waldmeier relations.
The 1D models have the much less cycle period than diffusive time
of the system. Therefore, we scale the periods of these models by
factor $\sim50$ . The Table 2 show the results for the mean and the
variance (standard deviations) for the period, amplitude, rise rate
and the shape of the sunspot cycles in the different data sets. From
that Table we see that the 1D1 model has the smaller variance in the
period, amplitude and rise rate of the cycles as compared to the others
data sets. The shape asymmetry of the cycles in 1D1 is opposite to
the others cases as well. Also we can see that the mechanism of the
helicity loss in the dynamo model influences the mean and variance
of the sunspot cycles parameters. In particular, the model 2D2 with
the increased diffusive loss of the magnetic helicity has the lower
variance of the period and amplitude of the sunspot cycles and has
the more symmetric shape of the cycle as compared to the model 2D1.
The difference in the synthetic data set of the sunspot cycles provided
by NIMV as compared with the SIDC is likely due to the fact that the
SIDC data set does not cover the periods with low magnetic activity.
This argument is also applied if we compared NIMV and, e.g., 2D1 model.
The parameters of the 2D1 model does not allow to have the extended
periods of time with very low sunspot cycles.

\begin{figure}[t]
 \includegraphics[width=0.9\textwidth]{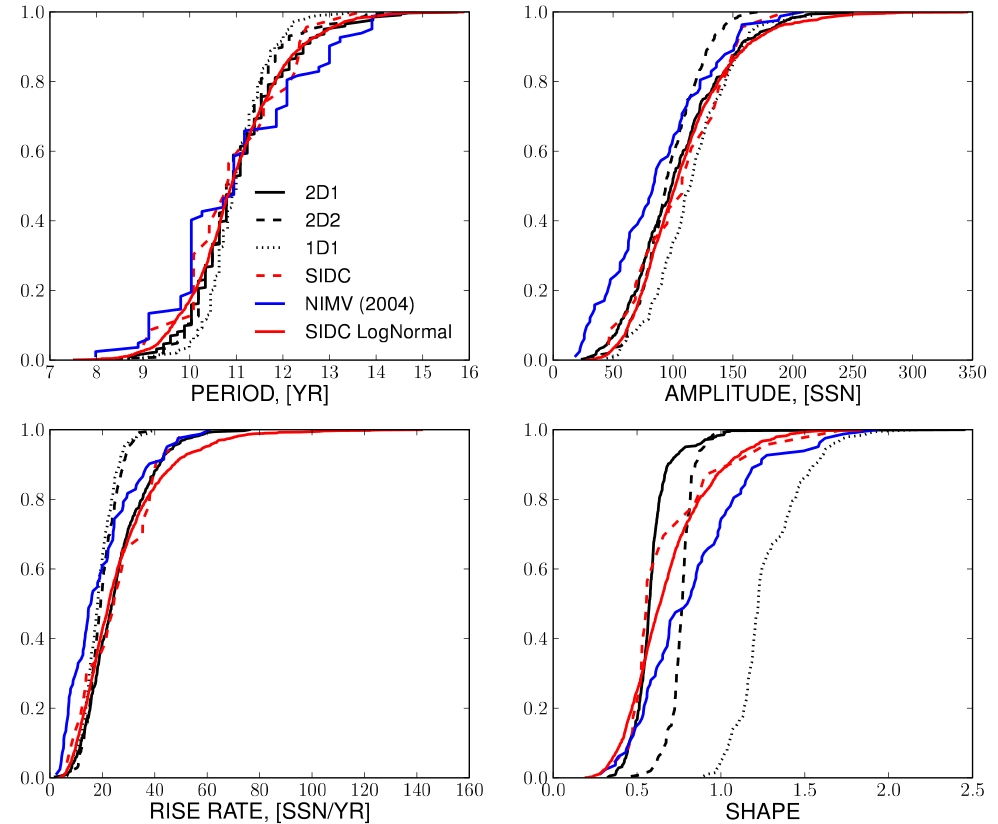} \caption{\label{fig:CDF}CDF distributions, red line - SIDC the data set, blue
line - the data set from \cite{nag04}.}
\end{figure}

The difference of the the statistical properties of the given data
set can be seen in further detail using the cumulative distribution
probability functions. The cumulative distributions are constructed
as follows. At the beginning, we sort each distribution for each parameter
and each model in increasing order. After this we compute the following
\begin{equation}
\mathrm{CDF}(P_{i})=\frac{\sum_{k=1}^{i}k}{N},\label{cdf}
\end{equation}
 where $P_{i}$ is the parameter under consideration (say, the cycle
period) having the order number $i$ (after sorting the set in
increasing order) and $N$ is the total number of the instances of
the given parameter in the set. Equation (\ref{cdf}) approximate the
probability for the parameter $P$ to have the values in interval
between $P_{\rm min}$ and $P_{i}$. The accuracy of the approximation
improves under $N\rightarrow\infty$. We will use the log-normal
cumulative distribution constructed on the base of the SIDC data set
as the reference distribution. The SIDC data set has only 23
instances of the sunspot cycles. To construct the reference
log-normal distribution we use the standard mean and variance of the
cycles parameters (period, amplitude, rise rate and asymmetry) given
in the Table 2. Then take the natural logarithm of them and
construct the log-normal distribution of the length 1000 using those
mean and variance. { The results are shown on the Fig.~6.}

It is clearly seen that log-normal distribution is a good fit for
the distributions of the sunspot cycles period in the SIDC data set
and also for model 2D2. The difference of the SIDC data set from the
log-normal distribution is seen in the probabilities distributions
for the rise rates and the shape of the cycles. It is, however, unclear
if these differences is due to the limited data set of cycles covered
by SIDC. The data set produced by the models and the NIMV data set
can be equally well approximated by the log-normal distributions (with
different mean and variance). For the dynamo models, the difference
between the distributions computed by Eq.(\ref{cdf}) and the log-normal
approximations for them is less visible than for SIDC and NIMV sets.

\begin{figure}
\includegraphics[width=0.9\textwidth]{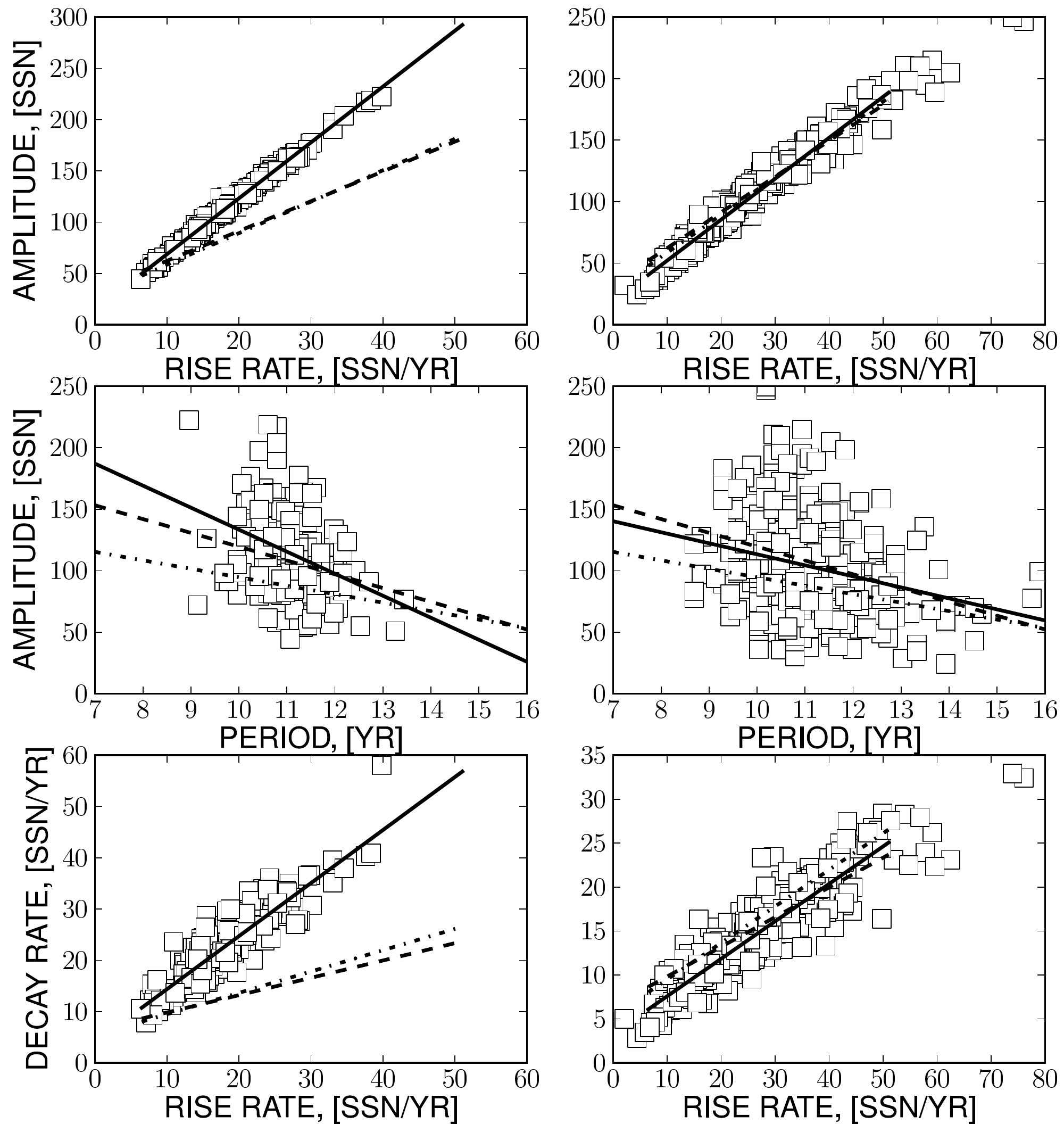} \caption{\label{fig:wald}The Waldmeier relations for 1D1 (left) and 2D1 (right)
models. The linear fits are shown the solid lines, the dashed lines
shows the fits for the SIDC data and the dash-dot line - for the NIMV
data.}
\end{figure}

Fig.~\ref{fig:wald} shows the Waldmeier relations for the 1D1 and
2D1 models together with their linear fits and also fits for the
SIDC and NIMV data sets. The parameters of the linear fits are
summarised in the Table 2. It is seen that the model 2D1 is well to
reproduce the SIDC data set, and the difference to the NIMV data is
not very large. The correspondence of the 2D2 model to the SIDC and
the NIMV is not as good as for the 2D1 model. This is also can be
expected by results presented in Fig.~\ref{fig:CDF} and by Table 2.
Finally, we can conclude that 1D1 model has only qualitative
agreement for the relations between the rise rate - amplitude, and
the period - amplitude of the sunspot cycles.

\section{Discussion and Conclusions}

In the paper we have studied the possibility to reproduce the statistical
relations of the sunspot activity cycle, like the so-called Waldmeier
relations, by means of the mean field dynamo model with the fluctuating
$\alpha$-effect. The dynamo model includes the long-term fluctuations
of the $\alpha$-effect. The dynamo models employ two types of the
nonlinear feedback of the mean-field on the $\alpha$-effect including
the algebraic quenching and the dynamic quenching due to the magnetic
helicity generation. The paper presents the results for three particular
dynamo models.

The presented 1D model is similar to model discussed by Moss et al.
\cite{moss-sok08}. It uses the constant shear and the algebraic
quenching of the $\alpha$-effect. The results for this model
disagree with observations (SIDC data set) about the shape of the
simulated sunspot (decay rate is higher than rise rate) even though
it is qualitatively reproduce the basic Waldmeier relations for the
Rise Rate-amplitude and the cycle Period-amplitude (see left column
in Fig.~\ref{fig:wald}). It was found that the variance of the cycle
parameters in the long-term evolution is less than in 2D models. It
is interesting, that under the level of noise the 1D models
involving the magnetic helicity show the smaller mean even though
having the stronger variances of the simulated sunspot parameters.
Although we could scale the mean parameters of those models to the
observational values, we did not present the results for these
models because they have the Waldmeier relations which are
quantitatively the same as those presented for 1D1 model in Table 2
and Fig.~\ref{fig:wald}.

We checked the 1D models with the spatially variable shear like that
suggested by Kitchatinov et al. \cite{kit-rud-kuk}. In agreement
with the helioseismology results, the given 1D models have the {
realistic latitudinal profile of the shear} (see
Eq.(\ref{eq:shear})). Although, these models qualitatively reproduce
the relation between the rise rate and amplitude of the cycle, they
fail with the other kind of relations, having the positive
correlation between the period and amplitude of the cycle and the
equal rate for the rise and decay phase of the simulated sunspot
cycles.

Similar to the 1D cases the magnetic helicity contribution to the
$\alpha$-effect results to decrease of the toroidal magnetic field
strength and to growth the variance of the cycle parameters in the
long-term evolution of the magnetic activity. The strong variance of
the cycle parameters is expected from SIDC data set and from NIMV as
well. For this reason in the paper we discuss the 2D model which
involves the effect of the magnetic helicity. The 2D models employ
two different description for the magnetic helicity loss, to
overcome the problem of the $\alpha$-effect catastrophic quenching.
The term $-\overline{\chi}/R_{\chi}\tau_{c}$ in Eq.(\ref{eq:hel})
describes the magnetic helicity loss with the dissipation rate
$(\tau_{c}R_{\chi})^{-1}$ without specifying the nature of the loss.
Note, that $\tau_{c}$ is varied from about 2 months near bottom of
the convection zone to a few hours at the top of the integration
domain (which is $0.99R_{\odot}$). Thus, for the $R_{\chi}=200$ used
in the model 2D1, the typical decay time for the magnetic helicity
is varied from about 4 solar cycles at the bottom of the convection
zone to a time which is less than one month at the top of the
convection zone. It is not clear if this simple description is
satisfactory approximation for the magnetic helicity loss. Therefore
we checked the alternative possibility using the diffusive helicity
flux. Although, the model that employ the diffusive helicity flux is
in satisfactory agreement with SIDC data, the correspondence to
observation in this model is not as good as for the model 2D1. We
find the the variance of the cycle parameters in the model 2D2 is
less than in the model 2D1 while the SIDC and NIMV data sets show
higher variances than the model 2D1.

The detailed comparison the results of our models with those given
by Karak and Choudhuri \cite{karak11} is not possible, because we
have used a different definition for the amplitude of the cycle and
the rise time. They did not give the results for the linear fits
coefficients and only provide the correlation coefficients in the
Waldmeier relations involving the Rise Rate-amplitude and the Rise
Time-Amplitude of the cycle. Bearing in mind the differences in
definition that their ``high diffusivity model'' with fluctuating
meridional circulation is comparable with our 2D1 and 2D2 models. It
is not clear however what is the typical shape of the cycle in their
model. This is an important issue as we have seen in example given
by model 1D1. It has qualitative agreement with SIDC data about the
period - amplitude and the rise rate - amplitude relations
even-though having the rise time of the cycle greater than the decay
time.

In the models under consideration, the asymmetry between the ascent
and decent phase of the sunspot cycle is inherent from the pattern
of the toroidal magnetic field activity. In particular, the 1D1 model
has the toroidal magnetic field butterfly diagram with maximum located
very close to equator. Therefore, applying the definition Eq.(\ref{eq:wolf})
for this type of the toroidal magnetic field evolutional pattern we
obtain the decent phase of the sunspot activity shorter than the
ascent phase. The opposite situation is in 2D models. There, we relate
the sunspot activity with the toroidal field in the subsurface layers.
The turbulent diffusivity in the model decrease outward this leads
to increases the decay time when the toroidal field gets closer to
the surface (see \cite{pk11}). We find that the effect of the magnetic
helicity on the $\alpha$-effect can amplify or saturate the asymmetry
of the cycle shape depending on the mechanisms of the helicity loss
employed in the model.

It is expected that the nonlinear dynamo mechanisms affect both the
magnetic cycle profile and the statistical properties of the cycles.
The paper illustrates the impact  of the non-linear $\alpha$-effect
for the algebraic and the dynamic non-linearities. Recently,
Kitchatinov \& Olemskoy \cite{kit-ole} suggested that the non-linear
diffusion could promotes the events similar to the Maunder minimum
provided there are the small fluctuations in the $\alpha$-effect.
This mechanism does not work in {\it our models}, because on the rise
phase of the cycle, the growing toroidal
 magnetic fields results to the turbulent diffusivity quenching
and this effect makes the rise phase of the cycle longer, i.e., the smaller
turbulent diffusion, the longer evolutionary time scale.
The opposite situation is expected for the decay phase of the magnetic cycle.

The comparison of the SIDC data set and the synthetic data set
provided by Nagovizyn et al. \cite{nag04}(NIMV) reveals the
significant difference in the statistical properties of the cycle
parameters. This seems to be a result of the wider cycle variations
range covering by the NIMV data set. The model presented in the
paper don't cover the variations seen in NIMV because the selected
models almost have no the extended events with low cycles like the
so-called Maunder minimum which were observed during the 16-th
century. This motivated us to extend our study and explore the
models which have the more intermittent variations of the sunspot
cycle. This work is planned for the future papers.

Summarizing the main findings of the paper we conclude as follows.
We found that the dynamo models,
having the reasonably good the time-latitude diagram of the toroidal
magnetic field evolution, are able to
reproduce qualitatively the inclination and dispersion across the
Waldmeier relations with less than 20\% Gaussian fluctuations of the
$\alpha$-effect. The 2D models have better agreement with
observations than 1D. In particular, 1D models fail to reproduce the
asymmetric shape of the sunspot cycle with short rise and long decay
phases. The statistical distributions of the cycle parameters show
the log-normal probability distributions for the all data sets
analysed in the paper. The parameters of these distributions are
different for all data sets. Again the 1D model is significantly
different from others in this sense. The 2D model that employs the
simplest form of the helicity loss via the term
$-\overline{\chi}/R_{\chi}\tau_{c}$ agrees well with the SIDC,
even-though the long-term variations in this model is not
intermittent enough, and this seems to be a reason for its
difference to the NIMV data set in some aspects. The employ of the
diffusive loss in the magnetic helicity evolution equation results
to decreasing in the variations of the cycle parameters. The further
study of the magnetic helicity transport mechanisms should clarify
the likely candidates which are responsible for the magnetic
helicity loss from the dynamo region. We have seen that the analysis
of the statistical relations of the sunspot cycle may provide the
valuable diagnostic tool for this study.
\section*{Acknowledgements}
The authors thanks the Nordita program "Dynamo, Dynamical Systems and
Topology" for the financial support. D.S. is grateful to RFBR for  
financial support under grant 09-05-00076-a and V.P. thanks 
for the financial support from NASA LWS NNX09AJ85G grant and for
 the partial support under RFBR grant 10-02-00148-a.

\section{Appendix }

We describe some parts of the mean-electromotive force. The basic
formulation is given in P08. For this paper we reformulate tensor
$\alpha_{i,j}^{(H)}$, which represents the hydrodynamical part of
the $\alpha$-effect, by using Eq.(23) from P08 in the following form,
\begin{eqnarray}
\alpha_{ij}^{(H)} & = & \delta_{ij}\left\{ 3\eta_{T}\left(f_{10}^{(a)}\left(\mathbf{e}\cdot\boldsymbol{\Lambda}^{(\rho)}\right)+f_{11}^{(a)}\left(\mathbf{e}\cdot\boldsymbol{\Lambda}^{(u)}\right)\right)\right\} +\label{eq:alpha}\\
 & + & e_{i}e_{j}\left\{ 3\eta_{T}\left(f_{5}^{(a)}\left(\mathbf{e}\cdot\boldsymbol{\Lambda}^{(\rho)}\right)+f_{4}^{(a)}\left(\mathbf{e}\cdot\boldsymbol{\Lambda}^{(u)}\right)\right)\right\} \nonumber \\
 & + & 3\eta_{T}\left\{ \left(e_{i}\Lambda_{j}^{(\rho)}+e_{j}\Lambda_{i}^{(\rho)}\right)f_{6}^{(a)}+\left(e_{i}\Lambda_{j}^{(u)}+e_{j}\Lambda_{i}^{(u)}\right)f_{8}^{(a)}\right\} .\nonumber
\end{eqnarray}
 The contribution of magnetic helicity $\overline{\chi}=\overline{\mathbf{a\cdot}\mathbf{b}}$
($\mathbf{a}$ is a fluctuating vector magnetic field potential) to
the $\alpha$-effect is defined as $\alpha_{ij}^{(M)}=C_{ij}^{(\chi)}\overline{\chi}$,
where
\begin{equation}
C_{ij}^{(\chi)}=2f_{2}^{(a)}\delta_{ij}\frac{\tau_{c}}{\mu_{0}\overline{\rho}\ell^{2}}-2f_{1}^{(a)}e_{i}e_{j}\frac{\tau_{c}}{\mu_{0}\overline{\rho}\ell^{2}}.\label{alpM}
\end{equation}
 The turbulent pumping, $\gamma_{i,j}$, is also part of the mean
electromotive force in Eq.(23)(P08). Here we rewrite it in a more
traditional form (cf, e.g., ),
\begin{equation}
\gamma_{ij}=3\eta_{T}\left\{ f_{3}^{(a)}\Lambda_{n}^{(\rho)}+f_{1}^{(a)}\left(\mathbf{e}\cdot\boldsymbol{\Lambda}^{(\rho)}\right)e_{n}\right\} \varepsilon_{inj}-3\eta_{T}f_{1}^{(a)}e_{j}\varepsilon_{inm}e_{n}\Lambda_{m}^{(\rho)},\label{eq:pump}
\end{equation}
 The effect of turbulent diffusivity, which is anisotropic due to
the Coriolis force, is given by:
\begin{equation}
\eta_{ijk}=3\eta_{T}\left\{ \left(2f_{1}^{(a)}-f_{1}^{(d)}\right)\varepsilon_{ijk}-2f_{1}^{(a)}e_{i}e_{n}\varepsilon_{njk}\right\} .\label{eq:diff}
\end{equation}
 Functions $f_{\{1-11\}}^{(a,d)}$ depend on the Coriolis number $\Omega^{*}=2\tau_{c}\Omega_{0}$
and the typical convective turnover time in the mixing-length approximation:
$\tau_{c}=\ell/u'$. They can be found in P08. The turbulent diffusivity
is parametrized in the form, $\eta_{T}=C_{\eta}\eta_{T}^{(0)}$, where
$\eta_{T}^{(0)}={\displaystyle \frac{u'\ell}{3}}$ is the characteristic
mixing-length turbulent diffusivity, $u'$ is the RMS convective velocity,
$\ell$ is the mixing length, $C_{\eta}$ is a constant to control
the intensity of turbulent mixing. The others quantities in Eqs.(\ref{eq:alpha},\ref{eq:pump},\ref{eq:diff})
are: $\mathbf{\boldsymbol{\Lambda}}^{(\rho)}=\boldsymbol{\nabla}\log\overline{\rho}$
is the density stratification scale, $\mathbf{\boldsymbol{\Lambda}}^{(u)}=\boldsymbol{\nabla}\log\left(\eta_{T}^{(0)}\right)$
is the scale of turbulent diffusivity, $\mathbf{e}=\boldsymbol{\Omega}/\left|\Omega\right|$
is a unit vector along the axis of rotation. Equations (\ref{eq:alpha},\ref{eq:pump},\ref{eq:diff})
take into account the influence of the fluctuating small-scale magnetic
fields, which can be present in the background turbulence and stem
from the small-scale dynamo (see discussions in). In our paper, the
parameter $\varepsilon={\displaystyle \frac{\overline{\mathbf{b}^{2}}}{\mu_{0}\overline{\rho}\overline{\mathbf{u}^{2}}}}$,
which measures the ratio between the magnetic and kinetic energies
of fluctuations in the background turbulence, is assumed equal to
1. This corresponds to the energy equipartition. The quenching function
of the hydrodynamical part of $\alpha$-effect is defined by
\begin{equation}
\psi_{\alpha}=\frac{5}{128\beta^{4}}\left(16\beta^{2}-3-3\left(4\beta^{2}-1\right)\frac{\arctan\left(2\beta\right)}{2\beta}\right).
\end{equation}
 Note, in notation of P08 $\psi_{\alpha}=-3/4\phi_{6}^{(a)}$, and
$\beta={\displaystyle \frac{\left|\overline{B}\right|}{u'\sqrt{\mu_{0}\overline{\rho}}}}$.
\end{document}